\documentclass{PoS}
\usepackage{amssymb,amsmath,epsfig}
\newcommand{\eq}[1]{Eq.~(\ref{#1})}

\newcommand{\be}{\begin{equation}}
\newcommand{\ee}{\end{equation}}
\newcommand{\bea}{\begin{eqnarray}}
\newcommand{\eea}{\end{eqnarray}}
\newcommand{\ben}{\begin{eqnarray*}}
\newcommand{\een}{\end{eqnarray*}}

\newcommand{\DS}{Dyson-Schwinger }

\newcommand{\w}{\omega}
\newcommand{\e}{\varepsilon}
\newcommand{\al}{\alpha}
\newcommand{\ba}{\beta}
\newcommand{\ga}{\gamma}
\newcommand{\G}{\Gamma}
\newcommand{\de}{\delta}

\newcommand{\si}{\sigma}

\newcommand{\ro}{\rho}

\newcommand{\ta}{\tau}

\newcommand{\ha}{\frac{1}{2}}
\newcommand{\pd}{\partial}

\renewcommand{\th}{\theta}

\newcommand{\cd}{{\cal D}}
\newcommand{\cs}{{\cal S}}
\newcommand{\cc}{{\cal C}}
\newcommand{\co}{{\cal O}}
\renewcommand{\div}{\vec{\nabla}}

\newcommand{\s}[2]{{#1}\!\cdot\!{#2}}
\newcommand{\ov}[1]{\overline{#1}}
\newcommand{\dk}[1]{\,\,\,\raisebox{-0.4ex}{\large $\bar{}$}\!\!d\,{#1}\,}

\newcommand{\ev}[1]{<\!\!{#1}\!\!>}

\title{Leading order QCD in Coulomb gauge}
\ShortTitle{Leading order QCD in Coulomb gauge}
\author{\speaker{Peter Watson}%
         \thanks{Work supported by the Deutsche Forschungsgemeinschaft 
(DFG) under contracts no. DFG-Re856/6-2,3.}\\
Institut f\"ur Theoretische Physik, Universit\"at T\"ubingen, 
Auf der Morgenstelle 14, D-72076 T\"ubingen, Deutschland\\
        E-mail: \email{watson@tphys.physik.uni-tuebingen.de}}
\author{Hugo Reinhardt\\
Institut f\"ur Theoretische Physik, Universit\"at T\"ubingen, 
Auf der Morgenstelle 14, D-72076 T\"ubingen, Deutschland
}
\abstract{Coulomb gauge QCD in the first order formalism can be written in terms of a ghost-free, nonlocal action that ensures total color charge conservation via Gauss' law.  Making an Ansatz whereby the nonlocal term (the Coulomb kernel) is replaced by its expectation value, the resulting Dyson-Schwinger equations can be derived.  With a leading order truncation, these equations reduce to the gap equations for the static gluon and quark propagators obtained from a quasi-particle approximation to the canonical Hamiltonian approach.  Moreover a connection to the heavy quark limit can be established, allowing an intuitive explanation for the charge constraint and infrared divergences.}
\FullConference{International Workshop on QCD Green's Functions, Confinement and Phenomenology\\
                 5-9 September 2011\\
                 Trento, Italy}
\begin{document}
\section{Introduction}

The \DS equations of Coulomb gauge quantum chromodynamics (QCD) represent one of the many techniques being currently explored in the hope of one day being able to describe confinement and the hadron spectrum from first principles.  As with most difficult problems, it is useful to be able to compare and contrast different approaches to gain further insight.  One aim of this talk is to compare a leading order truncation of the \DS equations \cite{arXiv:1111.6078} to the gap equations for the static gluon and quark propagators obtained within a quasi-particle approximation to the canonical Hamiltonian approach \cite{Szczepaniak:2001rg,Feuchter:2004mk,Adler:1984ri}.

The talk starts with a brief review of Coulomb gauge within the first order formalism, including a discussion of the charge constraint that emerges from the incompleteness of the gauge fixing.  To avoid problems stemming from the nonlocality of this formalism, an Ansatz is introduced such that the \DS equations can be derived.  The reduction of the truncated \DS equations to the gap equations for the static propagators and the link to heavy quarks will be shown.  How the heavy quark limit provides an intuitive explanation for the charge constraint and infrared divergences as being unobservable constant shifts in the potential will be discussed.

\section{Coulomb gauge in the first order formalism}

Let us begin by considering the functional integral associated with QCD (in Minkowski space):
\be
Z=\int\cd\Phi e^{\imath\cs_{QCD}},\;\;\cs_{QCD}=\int dx\left\{\ov{q}_{\al x}\left[\imath\ga^0D_{0x}+\imath\s{\vec{\ga}}{\vec{D}_{x}}-m\right]_{\al\ba}q_{\ba x}+\frac{1}{2}\s{\vec{E}_x^a}{\vec{E}_x^a}-\frac{1}{2}\s{\vec{B}_x^a}{\vec{B}_x^a}\right\}
\ee
where $\cd\Phi$ generically represents the functional integration measure over all fields present.  The (conjugate) quark field is ($\ov{q}$) $q_{\ba x}$ where the fundamental color, spin and flavor indices are denoted collectively with the index $\ba$ and the position argument with subscript $x$.  The Dirac $\ga$-matrices obey the usual Clifford algebra $\left\{\ga^\mu,\ga^\nu\right\}=2g^{\mu\nu}$ with metric $g^{\mu\nu}=\mbox{diag}(1,-\vec{1})$ (we explicitly extract all the minus signs associated with the metric such that all components of a spatial vector $\vec{x}$ are written with subscripts, i.e., $x_i$).  The temporal and spatial components of the covariant derivative in the fundamental color representation are given by
\be
D_{0x}=\pd_{0x}-\imath g\si_x^aT^a,\;\;\;\vec{D}_x=\div_x+\imath g\vec{A}_x^aT^a
\ee
where $\si_x^a$ ($=A_x^{0a}$) and $\vec{A}_x^a$ are the temporal and spatial components of the gluon field, respectively, and where the superscript $a$ denotes the color index in the adjoint representation.  The generators obey $\left[T^a,T^b\right]=\imath f^{abc}T^c$, where the $f^{abc}$ are the structure constants and we use the normalization $\mbox{Tr}[T^aT^b]=\de^{ab}/2$.  The chromoelectric and chromomagnetic fields are written in terms of the gluon field as
\be
\vec{E}_x^a=-\pd_{0x}\vec{A}_x^a-\vec{D}_x^{ab}\si_x^b,\;\;\;\;
\vec{B}_x^a=\div_x\times\vec{A}_x^a-\frac{1}{2}gf^{abc}\vec{A}_x^b\times\vec{A}_x^c
\ee
with the spatial component of the covariant derivative in the adjoint representation given by
\be
\vec{D}_x^{ab}=\de^{ab}\div_x-gf^{acb}\vec{A}_x^c.
\ee

The QCD action is invariant under gauge transforms $A\rightarrow A^\th=UAU^\dag-\imath/g(\pd U)U^\dag$, $q\rightarrow q^\th=Uq$, where $U_x=\exp{\{-\imath\th_x^aT^a\}}$ is a spacetime element of the $SU(N_c)$ group parametrized by $\th_x^a$.  Because of this invariance, the functional integral contains a divergence due to the integration over the gauge group.  When calculating Green's functions such as propagators, it is thus necessary to fix the gauge and our choice is Coulomb gauge: $\div\cdot\vec{A}=0$.  The Faddeev-Popov (FP) technique to fix the gauge involves inserting the identity
\be
1=\int\cd\th\de\left(F\left[\si^\th,\vec{A}^\th\right]\right)\mbox{Det}\left[M^{ab}(x,y)\right],\;\;\;\;
M^{ab}(x,y)=\left.\frac{\de F^a\left[\si_x^\th,\vec{A}_x^\th\right]}{\de\th_y^b}\right|_{F=0}
\label{eq:fp0}
\ee
into the functional integral.  However, in Coulomb gauge where $F=\div\cdot\vec{A}$ and the FP kernel reads $M(x,y)\sim-\div_x\cdot\vec{D}_x\de(x-y)$, there is an obvious problem when the gauge transform parameter $\th_x^a$ is spatially independent:
\be
-\div_x\cdot\vec{D}_x^{ab}\th^b(x_0)=0
\ee
(there are no temporal derivatives) such that the FP determinant automatically vanishes.  Coulomb gauge is incomplete in this respect.  The resolution of the temporal zero modes of the FP operator leads to a constraint on the total color charge of the system in the first order formalism \cite{Reinhardt:2008pr}.  The identity, \eq{eq:fp0}, is modified to
\be
1=\int\cd\ov{\th}\de\left(F\left[\si^\th,\vec{A}^\th\right]\right)\ov{\mbox{Det}}\left[M^{ab}(x,y)\right]
\label{eq:fp1}
\ee
where $\cd\ov{\th}$ and $\ov{\mbox{Det}}$ explicitly exclude the temporal zero modes, $\th(x_0)$.  The Coulomb gauge fixed functional integral is thus
\be
Z=\int\cd\Phi\de\left(\s{\div}{\vec{A}}\right)\ov{\mbox{Det}}\left[-\s{\div}{\vec{D}}\right]e^{\imath\cs_{QCD}}.
\ee
The conversion to the first order formalism goes as follows \cite{Zwanziger:1998ez,Watson:2006yq,Reinhardt:2008pr}.  An auxiliary vector field ($\vec{\pi}$) is introduced via
\be
\exp{\left\{\imath\int dx\frac{1}{2}\s{\vec{E}_x^a}{\vec{E}_x^a}\right\}}=\int\cd\pi\exp{\left\{\imath\int dx\left[-\frac{1}{2}\s{\vec{\pi}_x^a}{\vec{\pi}_x^a}-\s{\vec{\pi}_x^a}{\vec{E}_x^a}\right]\right\}}
\ee
and split up into components ($\phi$ is the longitudinal part of $\vec{\pi}$) with
\be
\mbox{const}=\int\cd\phi\cd\ta\exp{\left\{-\imath\int dx\,\ta_x^a\left(\s{\div_x}{\vec{\pi}_x^a}+\div_x^2\phi_x^a\right)\right\}}.
\ee
Changing variables $\vec{\pi}\rightarrow\vec{\pi}-\div\phi$ and integrating out the Lagrange multiplier, the functional integral now has the form
\be
Z=\int\cd\Phi\de\left(\s{\div}{\vec{A}}\right)\de\left(\s{\div}{\vec{\pi}}\right)\ov{\mbox{Det}}\left[-\s{\div}{\vec{D}}\right]
e^{\imath\cs'},
\ee
where the action, $\cs'$, is at most linear in the temporal gauge field, $\si$: the corresponding term is
\be
\cs_{\si}=\int dx\,\si_x^a\left(\s{\div_x}{\vec{D}_x^{ab}}\phi_x^b+gf^{abc}\s{\vec{A}_x^b}{\vec{\pi}_x^c}+g\ov{q}_{\al x}[\ga^0T^a]_{\al\ba}q_{\ba x}\right).
\ee
(Incidentally, the above form of the functional integral is the starting point for studying perturbation theory in the first order formalism \cite{Watson:2006yq,Watson:2007mz,Popovici:2008ty}.)  Importantly, the $\si$-field can be integrated out to give
\be
Z=
\int\cd\Phi\de\left(\s{\div}{\vec{A}}\right)\de\left(\s{\div}{\vec{\pi}}\right)
\ov{\mbox{Det}}\left[-\s{\div}{\vec{D}}\right]
\de\left(\s{\div}{\vec{D}}\phi+\hat{\ro}\right)
e^{\left(\imath\cs''\right)}
\ee
where the color charge, $\hat{\ro}$, includes both gluonic and quark contributions:
\be
\hat{\ro}_x^a=gf^{abc}\s{\vec{A}_x^b}{\vec{\pi}_x^c}+g\ov{q}_{\al x}[\ga^0T^a]_{\al\ba}q_{\ba x}.
\label{eq:ro0}
\ee
The $\phi$ field can be integrated out by using the eigenfunctions of the Faddeev-Popov operator as a complete orthonormal basis for an expansion, the crucial point being that one must remember the temporal zero modes \cite{Reinhardt:2008pr}.  Including the $\phi$-dependent part of the action, the explicit expression is
\be
\int\!\!\cd\phi\de\left(\s{\div}{\vec{D}}\phi+\hat{\ro}\!\right)
\exp{\left\{\!\frac{\imath}{2}\int\!dx\phi_x^a\div_x^2\phi_x^a\right\}}
=\de\left(\int d\vec{x}\hat{\ro}\!\right)
\ov{\mbox{Det}}\left[-\s{\div}{\vec{D}}\right]^{-1}
\!\!\exp{\left\{\!\!-\frac{\imath}{2}\int\!dx\hat{\ro}_x^a\hat{F}_x^{ab}\hat{\ro}_x^b\right\}}
\ee
where
\be
\hat{F}_x^{ab}=\left[-\s{\div_x}{\vec{D}_x^{ac}}\right]^{-1}\left(-\div_x^2\right)\left[-\s{\div_x}{\vec{D}_x^{cb}}\right]^{-1}.
\ee
Notice the appearance of the \emph{inverse} (modified) FP determinant, that will cancel against the original in the functional integral.  The $\de$-functional constraint that emerges constrains the total color charge, the spatial integral arising from the projection onto the temporal zero mode.  In order to study its effect, we rewrite this $\de$-functional constraint in Gaussian form \cite{arXiv:1111.6078}:
\be
\de\left(\int d\vec{x}\,\hat{\ro}\right)\sim\lim_{\cc\rightarrow\infty}{\cal N}(\cc)\exp{\left\{-\frac{\imath}{2}\int dx\,dy\,\hat{\ro}^a(x)\cc\de^{ab}\de(x_0-y_0)\hat{\ro}^b(y)\right\}}
\ee
where $\cc$ is a constant, ${\cal N}(\cc)$ is a normalization factor to be included implicitly in the functional integral measure, and the limit $\cc\rightarrow\infty$ will be taken only at the end of any calculation.  With this, our functional integral now reads
\be
Z=\int\cd\Phi\de\left(\s{\div}{\vec{A}}\right)\de\left(\s{\div}{\vec{\pi}}\right)e^{\imath\cs},
\label{eq:genfunc0}
\ee
with the action
\bea
\cs&=&\int dx\left\{\ov{q}_{\al x}\left[\imath\ga^0\pd_{0x}+\imath\s{\vec{\ga}}{\vec{D}_{x}}-m\right]_{\al\ba}q_{\ba x}-\ha\s{\vec{B}_x^a}{\vec{B}_x^a}-\ha\s{\vec{\pi}_x^a}{\vec{\pi}_x^a}+\s{\vec{\pi}_x^a}{\pd_{0x}\vec{A}_x^a}\right\}\nonumber\\&&
-\ha\int dx\,dy\hat{\ro}_x^a\tilde{F}^{ab}(x,y)\hat{\ro}_y^b
\label{eq:act0}
\eea
and where $\tilde{F}$ is the Coulomb kernel, but shifted by a spatial constant proportional to $\cc$:
\be
\tilde{F}^{ab}(x,y)=\hat{F}_x^{ab}\de(x-y)+\cc\de^{ab}\de(x_0-y_0).
\ee
There exists a useful connection between the Coulomb kernel and the temporal gluon propagator \cite{Cucchieri:2000hv}.  Redoing the analysis for the functional integral in the presence of a source ($\ro$) for the temporal gluon field, the temporal gluon propagator is defined as
\be
W_{\si\si}^{ab}(x,y)=\left.\frac{1}{Z[\ro]}\frac{\de^2Z[\ro]}{\de\imath\ro_x^a\de\imath\ro_y^b}\right|_{\ro=0}.
\ee
The presence of the source $\ro$ only alters the above action, \eq{eq:act0}, by replacing $\hat{\ro}$ with $\ov{\ro}=\hat{\ro}+\ro$.  As noted \cite{Cucchieri:2000hv}, the temporal gluon propagator has a purely instantaneous part given by the expectation value of the Coulomb kernel since it involves only spatial derivatives.  In our case, where the kernel is shifted by a constant, we see that
\be
W_{\si\si}^{ab}(x,y)\sim\ev{\imath F_x^{ab}\de(\vec{x}-\vec{y})+\imath\cc\de^{ab}}\de(x_0-y_0)+\mbox{non-inst.}
\ee

To recap, by writing the Coulomb gauge functional integral in the first order formalism, the FP determinant cancels after integrating out the temporal and longitudinal fields and Coulomb gauge is thus ghost-free \cite{Zwanziger:1998ez,Watson:2006yq}.  What remains of the gluon field are the two transverse vector components $\vec{A}$ and $\vec{\pi}$ (which would give rise to the two polarization states of photons in quantum electrodynamics).  Treating the temporal zero modes of the FP operator explicitly, it is further seen that the total color charge must be conserved and vanishing \cite{Reinhardt:2008pr}.  This is nothing more than the application of Gauss' law.  Writing the total charge constraint in Gaussian form, the Coulomb kernel is shifted by a spatial constant -- eventually however, we must take the limit where this constant diverges.  We shall see though that this is not a problem in the end.

\section{Truncated Dyson-Schwinger equations}

Having written down our functional integral in the first order formalism, we would like to use it.  Unfortunately, the Coulomb kernel term ($\tilde{F}$) is nonlocal because of the presence of the inverse FP operator.  In order to derive \DS equations, we therefore make a truncation Ansatz whereby we replace the Coulomb kernel with its expectation value \cite{arXiv:1111.6078}:
\be
\tilde{F}^{ab}(x,y)\rightarrow \left[F(\vec{x}-\vec{y})+\cc\right]\de^{ab}\de(x_0-y_0)
\ee
where $F$ is now some purely spatial, scalar function which will serve as nonperturbative input into the system.  Note that this Ansatz still includes the tree-level term, such that one-loop perturbative results could still be obtained at this stage.  The action is now local, and given the form of the color charge $\hat{\ro}$, \eq{eq:ro0}, the Coulomb interaction term $\hat{\ro}\tilde{F}\hat{\ro}$ now involves a set of effective four-point vertices (see below for their explicit form).  In effect, by converting to the first order formalism, we replace the dynamics of the nonperturbative towers of \DS equations \cite{Watson:2006yq} and Slavnov-Taylor identities \cite{Watson:2008fb} involving the temporal ($\si$), longitudinal ($\phi$) and ghost degrees of freedom with our leading order Ansatz for $F$.

Since we have only modified the Coulomb interaction part of the action, many of the propagator and vertex Green's functions in the present formalism can be read off from previous studies \cite{Watson:2006yq,Popovici:2008ty}.  In particular, the propagators ($W$ in our notation) in momentum space are given by:
\bea
W_{AAij}^{ab}(k)&=&\imath\de^{ab}t_{ij}(\vec{k})\frac{\G_{\pi\pi}(k)}{\Delta_g(k)},\nonumber\\
W_{A\pi ij}^{ab}(k)&=&-\de^{ab}k_0t_{ij}(\vec{k})\frac{\G_{A\pi}(k)}{\Delta_g(k)},\nonumber\\
W_{\pi\pi ij}^{ab}(k)&=&\imath\de^{ab}\vec{k}^2t_{ij}(\vec{k})\frac{\G_{AA}(k)}{\Delta_g(k)},\nonumber\\
W_{\ov{q}q\al\ba}(k)&=&-\frac{\imath}{\Delta_f(k)}\left[\ga^0k_0A_t(k)-\s{\vec{\ga}}{\vec{k}}A_s(k)+B_m(k)+\ga^0k_0\s{\vec{\ga}}{\vec{k}}A_d(k)\right]_{\al\ba},\nonumber\\
\Delta_g(k)&=&k_0^2\G_{A\pi}^2(k)-\vec{k}^2\G_{AA}(k)\G_{\pi\pi}(k)+\imath0_+,\nonumber\\
\Delta_f(k)&=&k_0^2A_t^2(k)-\vec{k}^2A_s^2(k)-B_m^2(k)+k_0^2\vec{k}^2A_d^2(k)+\imath0_+,
\label{eq:wgdecomp0}
\eea
where the various dressing functions arise from the decompositions of the proper two-point functions ($\G$):
\bea
\G_{\pi\pi ij}^{ab}(k)&=&\imath\de^{ab}\left[\de_{ij}\G_{\pi\pi}(k)+l_{ij}(\vec{k})\ov{\G}_{\pi\pi}(k)\right],\nonumber\\
\G_{A\pi ij}^{ab}(k)&=&\de^{ab}k_0\left[\de_{ij}\G_{A\pi}(k)+l_{ij}(\vec{k})\ov{\G}_{A\pi}(k)\right]=\G_{\pi Aij}^{ab}(-k),\nonumber\\
\G_{AAij}^{ab}(k)&=&\imath\de^{ab}\vec{k}^2\left[t_{ij}(\vec{k})\G_{AA}(k)+l_{ij}(\vec{k})\ov{\G}_{AA}(k)\right],\nonumber\\
\G_{\ov{q}q\al\ba}^{(0)}(k)&=&\imath\left[\ga^0k_0A_t(k)-\s{\vec{\ga}}{\vec{k}}A_s(k)-B_m(k)+\ga^0k_0\s{\vec{\ga}}{\vec{k}}A_d(k)\right]_{\al\ba}.
\label{eq:gdecomp0}
\eea
In the above, $l_{ij}$ and $t_{ij}$ are the usual longitudinal and transverse spatial projectors, respectively.  The components of the gluon propagator are spatially transverse because we are in Coulomb gauge.  The dressing functions are all scalar functions of $k_0^2$ and $\vec{k}^2$ separately, due to the noncovariance.  At tree-level $\G_{\pi\pi}=\G_{A\pi}=\G_{AA}=A_t=A_s=1$, $B_m=m$ and all others vanish.  Notice the matrix inversion structure of the components of the gluonic and quark propagators, with the denominator factors $\Delta_g$ and $\Delta_f$ -- these will turn out to be important.  The tree-level quark-gluon ($\G_{\ov{q}qA}$), three- ($\G_{AAA}$) and four-gluon ($\G_{AAAA}$) vertices are also unaltered from \cite{Watson:2006yq,Popovici:2008ty}, although their explicit form will not be needed here.  With our Ansatz to replace the Coulomb kernel with its expectation value, the new tree-level vertices explicitly read \cite{arXiv:1111.6078} (all momenta incoming)
\begin{align}
\G_{AA\pi\pi ijkl}^{(0)abcd}(k_1,k_2,k_3,k_4)&=
-\imath g^2\left[f^{ead}f^{fbc}\de_{il}\de_{jk}\tilde{F}^{ef}(k_1\!+\!k_4)
+f^{ebd}f^{fac}\de_{jl}\de_{ik}\tilde{F}^{ef}(k_1\!+\!k_3)\right],\nonumber\\
\G_{\ov{q}qA\pi\al\ba ij}^{(0)ab}(k_1,k_2,k_3,k_4)&=
\imath g^2f^{abe}\!\left[\!\ga^0T^f\!\right]_{\al\ba}\!\de_{ij}\tilde{F}^{ef}(k_1\!+\!k_2),\nonumber\\
\G_{\ov{q}q\ov{q}q\al\ba\ga\de}^{(0)}(k_1,k_2,k_3,k_4)&=
-\imath g^2\!\left[\!\ga^0T^a\!\right]_{\al\ba}\!\left[\!\ga^0T^b\!\right]_{\ga\de}\!\tilde{F}^{ab}(k_1\!+\!k_2)
+\imath g^2\!\left[\!\ga^0T^a\!\right]_{\al\de}\!\left[\!\ga^0T^b\!\right]_{\ga\ba}\!\tilde{F}^{ba}(k_1\!+\!k_4).
\label{eq:treev1}
\end{align}

\begin{figure}[t]
\vspace{0.8cm}
\begin{center}
\includegraphics[width=0.7\linewidth]{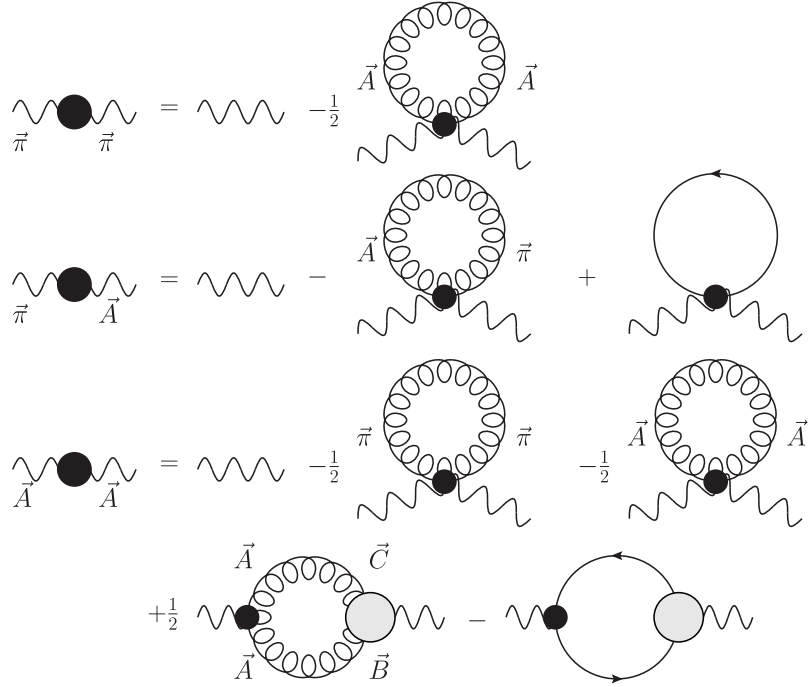}
\end{center}
\vspace{0.3cm}
\caption{\label{fig:gdse0}\DS equations for $\G_{\pi\pi}$, $\G_{\pi A}$ and $\G_{AA}$, omitting two-loop terms.  Wavy lines denote proper functions, the large filled blob indicating the dressed function.  Springs denote gluonic propagators, lines denote the quark propagator and all internal propagators are dressed.  Small blobs indicate tree-level vertices and large circles denote dressed vertices.  The gluonic field types $\vec{B}$ and $\vec{C}$ denote the sum over $\vec{A}$ and $\vec{\pi}$ contributions arising due to the presence of mixed gluon propagators.  See text for details.}
\end{figure}

\begin{figure}[t]
\vspace{0.8cm}
\begin{center}
\includegraphics[width=0.6\linewidth]{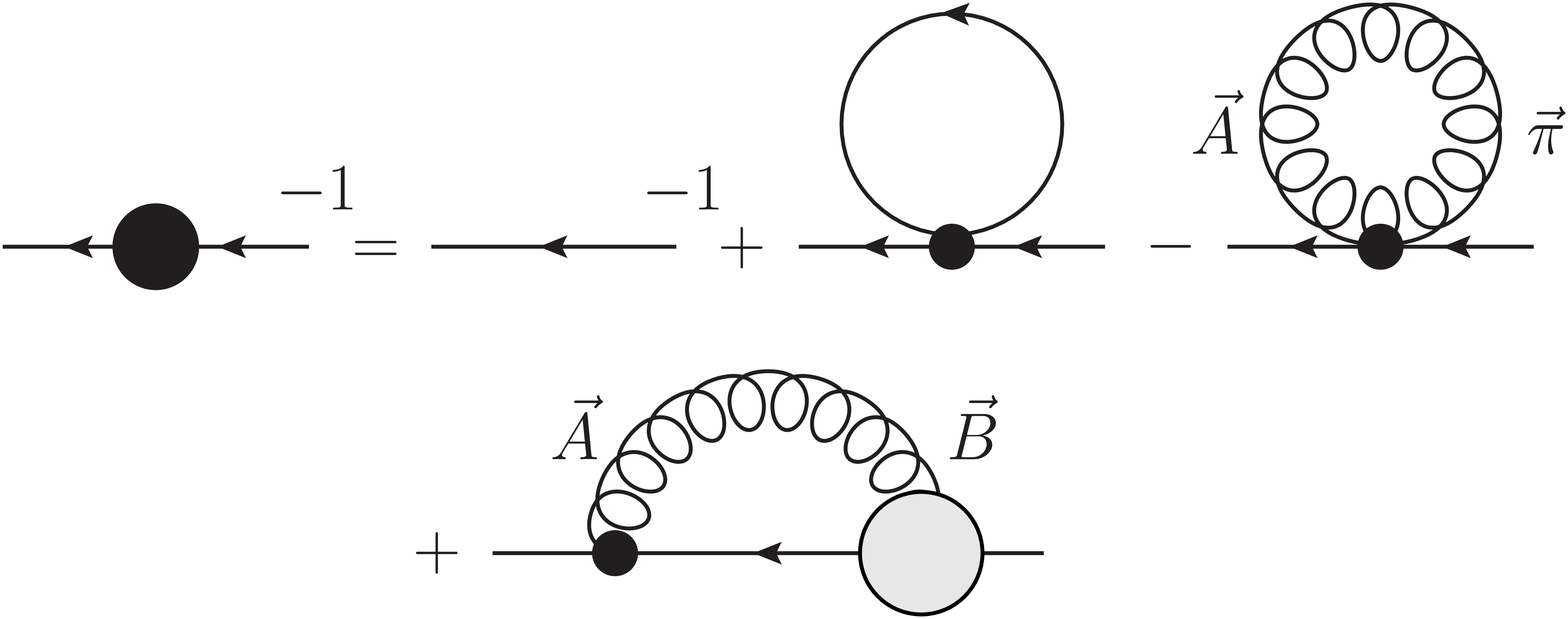}
\end{center}
\vspace{0.3cm}
\caption{\label{fig:qdse0}\DS equation for the quark two-point function, omitting two-loop terms.  On the left-hand side, the filled blob indicates the dressed (inverse) propagator, otherwise notation is as in the previous figure.  See text for details.}
\end{figure}

With a little practice, the \DS equations are not difficult to derive (although keeping track of the signs when quarks are present is somewhat tedious).  Generically, their structure arises from the Legendre transform and repeated functional differentiation of the generating functional, giving the characteristic sequence of loop integrals.  Such a derivation in Coulomb gauge is given in Refs.~\cite{Watson:2006yq,Watson:2007vc,Popovici:2008ty}.  Omitting the two-loop contributions, the \DS equations for the proper two-point functions, in the system considered here \cite{arXiv:1111.6078}, are presented diagrammatically in Figs.~\ref{fig:gdse0} and \ref{fig:qdse0}.
Because of the existence of the mixed gluon propagator $W_{A\pi}$, certain loops involve a sum over the two gluonic field types $\vec{A}$, $\vec{\pi}$ which is denoted by $\vec{B}$ and $\vec{C}$ in the diagrams.  In addition to the truncation to omit two-loop contributions, we further restrict to considering only those terms arising from the tree-level four-point vertices involving the Coulomb kernel $\tilde{F}$, i.e., we throw away the $\G_{\ov{q}qA}$, $\G_{AAA}$ and $\G_{AAAA}$ tree-level vertices.  The remaining loops of the \DS equations are thus tadpole contributions involving the propagators and our input Ansatz for $\tilde{F}$, forming a closed set of equations.  The input we have in mind is motivated by the connection to the instantaneous part of the temporal gluon propagator.  In momentum space and omitting the perturbative contributions, we will assume the strongly infrared enhanced form:
\be
g^2C_F\tilde{F}^{ab}(k)=\de^{ab}F(\vec{k}^2)+\de^{ab}\cc(2\pi)^3\de(\vec{k}),\;\;\;\;F(\vec{k}^2)=8\pi\si_c/\vec{k}^4
\label{eq:ansf0}
\ee
where $C_F=(N_c^2-1)/2N_c$ and $\si_c$ is the Coulomb string tension \cite{Zwanziger:2002sh}.  Note that $g^2\tilde{F}$ is a renormalization group invariant quantity in Coulomb gauge \cite{Zwanziger:1998ez,Cucchieri:2000hv}, so is ideal for use as input.  We shall show that this truncation results in gap equations for the static gluon and quark propagator dressing functions that can be compared to those derived in the canonical Hamiltonian approach, Refs.~\cite{Szczepaniak:2001rg} and \cite{Adler:1984ri}, respectively.

\section{Leading order static gluon equation}

Let us consider the truncated \DS equation for the mixed gluonic proper two-point function, $\G_{\pi A}$ (middle line of Fig.~\ref{fig:gdse0}).  Recognizing that the color structure of the quark tadpole loop vanishes, the equation can be written ($\dk{\w}=d^4\w/(2\pi)^4$)
\be
\de_{ij}\G_{A\pi}(k)+l_{ij}(\vec{k})\ov{\G}_{A\pi}(k)=\de_{ij}
-\imath g^2N_c\int\frac{\dk{\w}\w_0\G_{A\pi}(\w)}{k_0\Delta_g(\w)}t_{ij}(\vec{\w})\tilde{F}(k-\w)
\ee
where we have expanded the two-point functions using Eqs.~(\ref{eq:wgdecomp0},\ref{eq:gdecomp0}) and the four-point function, \eq{eq:treev1}, subsequently resolving the color structure.  Since $\tilde{F}$ is energy independent (coming from the instantaneous Coulomb kernel) and all dressing functions are even functions of energy, the energy integral of the above is overall odd and vanishes.  This immediately gives the results
\be
\G_{A\pi}=1,\;\;\;\;\ov{\G}_{A\pi}=0,\;\;\;\;\Delta_g(k)=k_0^2-\vec{k}^2\G_{AA}(k)\G_{\pi\pi}(k)+\imath0_+.
\label{eq:entriv}
\ee
Turning to the truncated \DS equations for $\G_{\pi\pi}$ and $\G_{AA}$ (first and last lines of Fig.~\ref{fig:gdse0}, respectively), after sorting out the decompositions and color factors as above, we have (the longitudinal dressing functions $\ov{\G}_{\pi\pi}$ and $\ov{\G}_{AA}$ play no role here)
\bea
\G_{\pi\pi}(k)&=&1
+\frac{\imath}{2}g^2N_c\int\frac{\dk{\w}\G_{\pi\pi}(\w)}{\Delta_g(\w)}\tilde{F}(k-\w)t_{ji}(\vec{k})t_{ij}(\vec{\w}),\nonumber\\
\G_{AA}(k)&=&1
+\frac{\imath}{2}g^2N_c\int\frac{\dk{\w}\vec{\w}^2\G_{AA}(\w)}{\vec{k}^2\Delta_g(\w)}\tilde{F}(k-\w)t_{ji}(\vec{k})t_{ij}(\vec{\w}).
\eea
The energy integrals do not involve $k_0$ because of the energy independence of $\tilde{F}$ and we thus see that $\G_{\pi\pi}$ and $\G_{AA}$ are energy independent.  The energy dependence of the denominator factor $\Delta_g$ is now reduced such that we can now define the static (i.e., energy integrated or equaltime) gluon propagator in terms of a single dressing function, $G$ \cite{arXiv:1111.6078}:
\be
W_{AAij}^{(s)ab}(\vec{k})=\int\frac{dk_0}{2\pi}W_{AAij}^{ab}(k)=\de^{ab}t_{ij}(\vec{k})\frac{1}{2|\vec{k}|}G(\vec{k}^2)^{1/2},\;\;\;\;G=\G_{\pi\pi}/\G_{AA}.
\label{eq:gdef0}
\ee
A similar expression exists for the static $\pi$-propagator, $W_{\pi\pi}^{(s)}$.  Further inserting the Ansatz form for $\tilde{F}$ from \eq{eq:ansf0} and writing the spatial integrals in terms of $G$, the equations become
\bea
\G_{\pi\pi}(\vec{k}^2)&=&1+\frac{N_c}{2C_F}\frac{{\cal C}}{\sqrt{\vec{k}^2}}G(\vec{k}^2)^{1/2}
+\frac{N_c}{4C_F}\int\frac{\dk{\vec{\w}}}{\sqrt{\vec{\w}^2}}G(\vec{\w}^2)^{1/2}F(\vec{k}-\vec{\w})t_{ji}(\vec{k})t_{ij}(\vec{\w}),\nonumber\\
\G_{AA}(\vec{k}^2)&=&1+\frac{N_c}{2C_F}\frac{{\cal C}}{\sqrt{\vec{k}^2}}G(\vec{k}^2)^{-1/2}
+\frac{N_c}{4C_F}\int\frac{\dk{\vec{\w}}}{\sqrt{\vec{\w}^2}}\frac{\vec{\w}^2}{\vec{k}^2}G(\vec{\w}^2)^{-1/2}F(\vec{k}-\vec{\w})t_{ji}(\vec{k})t_{ij}(\vec{\w})
\eea
where $\dk{\vec{\w}}=d\vec{\w}/(2\pi)^3$.  The proper dressing functions $\G_{\pi\pi}$ and $\G_{AA}$ have contributions linear in $\cc$ (the constant that arises from the charge conservation) and also from the potentially infrared divergent spatial integrals over $F\sim1/(\vec{k}-\vec{\w})^4$, if we use \eq{eq:ansf0} as input.  However, further utilizing the definition of $G$, \eq{eq:gdef0}, we find that we can combine the above coupled equations into a single equation for the static gluon propagator dressing function:
\be
G(\vec{k}^2)=1+\frac{1}{4}\frac{N_c}{C_F}\int\frac{\dk{\vec{\w}}}{\sqrt{\vec{\w}^2}}F(\vec{k}-\vec{\w})t_{ji}(\vec{k})t_{ij}(\vec{\w})\left[G(\vec{\w}^2)^{1/2}-\frac{\vec{\w}^2}{\vec{k}^2}\frac{G(\vec{k}^2)}{G(\vec{\w}^2)^{1/2}}\right].
\label{eq:ggap0}
\ee
This is the gluon gap equation and is identical to that originally derived from the canonical approach \cite{Szczepaniak:2001rg}.  The troublesome terms proportional to $\cc$ drop out and the infrared divergence of the spatial integrals is canceled (this is explicitly verified in \cite{arXiv:1111.6078}).  It thus appears that under this (leading order) truncation, the static gluon propagator contains the physical dynamics of the system whereas the full propagator (in particular, its pole position) is unphysical.  We shall discuss this at the end of the next section.  It is known that for an interaction of the type given by \eq{eq:ansf0}, the solution to \eq{eq:ggap0} is of the massive type \cite{arXiv:1111.6078,Szczepaniak:2001rg}, in contradiction to the expected Gribov type solution \cite{Gribov:1977wm}.  However, from the canonical approach, it is known that the gap equation receives significant infrared contributions from the ghost loop (`curvature') \cite{Feuchter:2004mk} which is missing from the leading order truncation presented here.

\section{Leading order quarks and the heavy limit}

The analysis for the quark \DS equation is very similar to that previously described for the gluon.  This similarity arises because the color charge $\hat{\ro}$, \eq{eq:ro0}, treats the gluonic and quark contributions on an equal footing.  Truncating the equation (Fig.~\ref{fig:qdse0}), inserting the appropriate factors, Eqs.~(\ref{eq:wgdecomp0},\ref{eq:gdecomp0},\ref{eq:treev1}), resolving the color factors and projecting out the Dirac components, we obtain four coupled equations (one for each of the dressing functions).  Two are trivial because they involve odd energy integrals:
\bea
A_t(k)&=&1-\imath g^2C_F\int\frac{\dk{\w}\w_0A_t(\w)\tilde{F}(k-\w)}{k_0\Delta_f(\w)},\nonumber\\
A_d(k)&=&\imath g^2C_F\int\frac{\dk{\w}\w_0\s{\vec{k}}{\vec{\w}}A_d(\w)\tilde{F}(k-\w)}{k_0\vec{k}^2\Delta_f(\w)},
\eea
such that
\be
A_t=1,\;\;\;\;A_d=0,\;\;\;\;\Delta_f(k)=k_0^2-\vec{k}^2A_s^2(k)-B_m^2(k)+\imath0_+.
\ee
The other two equations are
\bea
A_s(k)&=&1+\imath g^2C_F\int\frac{\dk{\w}\s{\vec{k}}{\vec{\w}}A_s(\w)\tilde{F}(k-\w)}{\vec{k}^2\Delta_f(\w)},\nonumber\\
B_m(k)&=&m+\imath g^2C_F\int\frac{\dk{\w}B_m(\w)\tilde{F}(k-\w)}{\Delta_f(\w)},
\eea
where the energy independence of $\tilde{F}$ again means that $A_s$ and $B_m$ are purely spatial.  As for the gluon, we can now write the static quark propagator in terms of a single dressing function, $M$:
\be
W_{\ov{q}q\al\ba}^{(s)}(\vec{k})=\int\frac{dk_0}{2\pi}W_{\ov{q}q\al\ba}(k)=\frac{\left[\s{\vec{\ga}}{\vec{k}}-M(\vec{k}^2)\right]_{\al\ba}}{2\sqrt{\vec{k}^2+M(\vec{k}^2)^2}},\;\;\;\;M(\vec{k}^2)=\frac{B_m(\vec{k}^2)}{A_s(\vec{k}^2)}.
\ee
Inserting the form of $\tilde{F}$ given by \eq{eq:ansf0}, these equations become
\bea
A_s(\vec{k}^2)&=&1+\frac{1}{2}\frac{{\cal C}}{\sqrt{\vec{k}^2+M(\vec{k}^2)^2}}+\frac{1}{2}\int\frac{\dk{\vec{\w}}\s{\vec{k}}{\vec{\w}}F(\vec{k}-\vec{\w})}{\vec{k}^2\sqrt{\vec{\w}^2+M(\vec{\w}^2)^2}},\nonumber\\
B_m(\vec{k}^2)&=&m+\frac{1}{2}\frac{{\cal C}M(\vec{k}^2)}{\sqrt{\vec{k}^2+M(\vec{k}^2)^2}}+\frac{1}{2}\int\frac{\dk{\vec{\w}}M(\vec{\w}^2)F(\vec{k}-\vec{\w})}{\sqrt{\vec{\w}^2+M(\vec{\w}^2)^2}}.
\label{eq:dqgap0}
\eea
Like for the gluon, the dressing functions for the quark propagator are also linear in $\cc$ and involve potentially infrared divergent spatial integrals.  However, the coupled equations can also be combined into a single gap equation for $M$:
\be
M(\vec{k}^2)=m+\frac{1}{2}\int\frac{\dk{\vec{\w}}F(\vec{k}-\vec{\w})}{\sqrt{\vec{\w}^2+M(\vec{\w}^2)^2}}\left[M(\vec{\w}^2)-\frac{\s{\vec{k}}{\vec{\w}}}{\vec{k}^2}M(\vec{k}^2)\right]
\label{eq:qgap0}
\ee
where the $\cc$-dependence and the potential infrared divergence cancels (also verified explicitly in Ref.~\cite{arXiv:1111.6078}).  This equation is very well-known as the Adler-Davis truncation \cite{Adler:1984ri} and was originally derived using the canonical approach.  With the interaction, \eq{eq:ansf0}, the solution does exhibit dynamical chiral symmetry breaking \cite{arXiv:1111.6078,Adler:1984ri}, but to a quantitatively too small degree for phenomenology -- this leading order truncation requires further contributions \cite{Pak:2011wu}.

It is possible to make the connection between the truncated quark propagator and the known Coulomb gauge heavy quark limit \cite{Popovici:2010mb} (see also Carina Popovici's contribution to these proceedings).  This is done via a spin-decomposition of the full quark propagator \cite{Eichten:1980mw}.  Introducing the spin projectors
\be
P_\pm=(1\pm\ga^0)/2,\;\;P_++P_-=1
\ee
the full quark propagator can be written as
\be
W_{\ov{q}q\al\ba}(k)=\left[(P_++P_-)W_{\ov{q}q}(k)(P_++P_-)\right]_{\al\ba}.
\ee
We now consider the heavy quark limit in the Coulomb gauge rest frame: $|\vec{k}|/m\rightarrow0$.  Using \eq{eq:qgap0}, we can make an estimate for the static dressing function $M$ (this is confirmed numerically in \cite{arXiv:1111.6078}).  The function $F(\vec{k}-\vec{\w})$ peaks at $\vec{\w}=\vec{k}$ but the bracketed combination of functions vanishes, canceling the infrared divergence and leaving
\be
M(\vec{k}^2)\approx m+\#\frac{M(\vec{k}^2)}{\sqrt{\vec{k}^2+M^2(\vec{k}^2)}}\stackrel{|\vec{k}|\ll m}{\rightarrow}m+\#
\ee
so that from \eq{eq:dqgap0}, the functions $A_s$ and $B_m$ are given by
\bea
A_s(\vec{k}^2)&\stackrel{|\vec{k}|\ll m}{\rightarrow}&1+\co(1/m)\nonumber\\
B_m(\vec{k}^2)&\stackrel{|\vec{k}|\ll m}{\rightarrow}&m+\frac{1}{2}\cc+\frac{1}{2}\int\dk{\vec{\w}}F(\vec{\w}^2)=B_h
\eea
where for $B_m$ it is recognized that the infrared divergence of the integral is not suppressed by factors of $1/m$, and for $A_s$ we
demand that $m$ is the largest scale (the limit $\cc\rightarrow\infty$ is taken only at the end).  The spin-decomposed quark propagator is then
\bea
W_{\ov{q}q\al\ba}(k)&\stackrel{|\vec{k}|\ll m}{\rightarrow}&\frac{(-\imath)}{[k_0^2-B_h^2+\imath0_+]}\left\{[k_0+B_h]P_+P_+-[k_0-B_h]P_-P_-\right\}_{\al\ba}\nonumber\\
&=&-\imath\frac{[P_+P_+]_{\al\ba}}{[k_0-B_h+\imath\e]}+\imath\frac{[P_-P_-]_{\al\ba}}{[k_0+B_h-\imath\e]}.
\eea
The first component represents a heavy quark propagating forward in time (the second is the antiquark propagating backwards in time) and explicitly agrees with the expression found in Ref.~\cite{Popovici:2010mb}.  Further, when considering the heavy quark limit of the Bethe-Salpeter equation, the Faddeev equation or the quark four-point function, the constant $\cc$ and infrared divergent spatial integral occurring in $B_h$ cancel explicitly when the quarks are in a color singlet configuration (and only for these configurations).  The Bethe-Salpeter equation moreover furnishes the result that
\be
V(r)\sim\int\dk{\vec{\w}}F(\vec{\w}^2)(1-e^{\imath\vec{\w}\cdot\vec{r}})
\ee
($r$ is a length scale) so that the connection between the Coulomb kernel, $F$, and the quark-antiquark potential, $V$, is made explicit.

The connection to the heavy quark limit is rather important, because it explains the role of the constant $\cc$ (arising from the charge constraint) and the infrared divergences.  Within the leading order truncation presented here, the physical dynamics are contained within the static propagator dressing functions and their gap equations.  The full propagators have pole positions that are dependent on $\cc$ and the infrared divergence.  As the pole positions are shifted to infinity in the limit $\cc\rightarrow\infty$ (i.e., when the total color charge is conserved and vanishing) and as the infrared integrals diverge, this simply reflects the fact that infinite energy is required to create isolated colored particles from the colorless vacuum.  However, for physical color singlets the divergent contributions cancel.  It thus appears that the constant $\cc$ and the infrared divergence are merely constant shifts in the potential and which are not observable.

\section{Summary}

In summary, a leading order truncation to the \DS equations of Coulomb gauge within the first order formalism has been presented.  Because Coulomb gauge is incomplete, the temporal zero modes must be taken into account and it is seen that this results in a nonperturbative constraint on the total color charge.  In the Coulomb gauge first order formalism, the ghosts cancel but the resulting action is nonlocal.  To derive the \DS equations, an Ansatz is thus made to replace the nonlocal Coulomb kernel with its expectation value.  This introduces a set of four-point interaction vertices, which are dependent on the input Ansatz for the Coulomb kernel ($F$).  Truncating the system to include only the tadpole diagrams involving this input Ansatz leads to a closed set of equations (their solution is discussed in Ref.~\cite{arXiv:1111.6078}).  Importantly, these equations reduce to the gap equations for the static gluon and quark propagators obtained from a quasi-particle approximation in the canonical Hamiltonian approach, Refs.~\cite{Szczepaniak:2001rg} and \cite{Adler:1984ri}, respectively.  Furthermore, the known Coulomb gauge heavy quark limit \cite{Popovici:2010mb} emerges.  It is seen that the static propagator gap equations are not affected by the charge constraint or the infrared divergence of the input Ansatz, $F$.  This is in contrast to the full propagators, which are dependent.  However, the connection to the heavy quark limit supplies an explanation: such unphysical singularities cancel for color singlet states, whereas the pole positions of colored propagators are shifted to infinity, reflecting that infinite energy is required for these to be created in isolation from the vacuum.

\begin{acknowledgments}
It is a pleasure to thank the organizers (Daniele Binosi in particular) for a most enjoyable and informative workshop.
\end{acknowledgments}


\end{document}